\documentclass[lettersize,journal]{IEEEtran}
\usepackage{amsmath,amsfonts}
\usepackage{algorithmic}
\usepackage{algorithm}
\usepackage{array}
\usepackage{comment}
\usepackage{upgreek}
\usepackage{MnSymbol}
\usepackage{textcomp}
\usepackage{hyperref}

\usepackage{newtxtext,newtxmath}
\usepackage[caption=false,font=normalsize,labelfont=rm,textfont=rm,position=top]{subfig}

\usepackage{setspace}

\usepackage{ulem}
\usepackage{stfloats}
\usepackage{url}
\usepackage{verbatim}
\usepackage{graphicx}
\usepackage{cite}
\begin{document}

\title{Next Generation Ta-STJ Sensor Arrays for BSM Physics Searches}



\author{
  J.~P.~T.~Templet\IEEEauthorrefmark{1}\IEEEauthorrefmark{2},~%
  S.~Fretwell\IEEEauthorrefmark{1},~%
  A.~Marino\IEEEauthorrefmark{1},~%
  R.~Cantor\IEEEauthorrefmark{3},~%
  A.~Hall\IEEEauthorrefmark{3},~%
  C.~Bray\IEEEauthorrefmark{1},~%
  \mbox{C.~Stone-Whitehead\IEEEauthorrefmark{1},}\\%
  I.~Kim\IEEEauthorrefmark{4},~%
  F.~Ponce\IEEEauthorrefmark{5},~%
  W.~Van~De~Pontseele\IEEEauthorrefmark{1},~%
K.~G.~Leach\IEEEauthorrefmark{1}\IEEEauthorrefmark{6},~
  S.~Friedrich\IEEEauthorrefmark{4}~%
  \thanks{\IEEEauthorrefmark{1}Department of Physics, Colorado School of Mines, Golden, CO, 80401, USA}%
  \thanks{\IEEEauthorrefmark{2}(E-mail: \href{mailto:josephtemplet@mines.edu}{josephtemplet@mines.edu})}
  \thanks{\IEEEauthorrefmark{3}STAR Cryoelectronics LLC, Santa Fe, NM 87508, USA.}%
    \thanks{\IEEEauthorrefmark{4}Lawrence Livermore National Laboratory, Livermore, CA 94550, USA.}%
    \thanks{\IEEEauthorrefmark{5}Pacific Northwest National Laboratory, Richland, WA, 99354, USA.}%
    \thanks{\IEEEauthorrefmark{6}Facility for Rare Isotope Beams, Michigan State University, East Lansing, MI, 48824, USA.}%
\thanks{Manuscript Received September 26, 2025}%
\thanks{Revisions Received January 16, 2026}%
}

\markboth{IEEE TRANSACTIONS ON APPLIED SUPERCONDUCTIVITY, ~Vol.~??, No.~??, October~2025}%
{Templet, Author2, \MakeLowercase{\textit{et al.}}}

\IEEEpubid{0000--0000/00\$00.00~\copyright~2025 IEEE}
\begin{spacing}{1}
\maketitle

\begin{abstract}

The Beryllium Electron capture in Superconducting Tunnel junctions (BeEST) experiment uses superconducting tunnel junction (STJ) sensors to search for physics beyond the standard model (BSM) with recoil spectroscopy of the $\mathbf{^7}$Be EC decay into $\mathbf{^7}$Li. A pulsed UV laser is used to calibrate the STJs throughout the experiment with $\sim$20 meV precision. Phase-III of the BeEST experiment revealed a systematic calibration discrepancy between STJs. We found these artifacts to be caused by resistive crosstalk and by variable substrate heating due to intensity variations of the calibration laser. For phase-IV of the BeEST experiment, we have removed the crosstalk by designing the STJ array so that each pixel has its own ground wire. We now also use a more stable UV laser for calibration. The new STJ arrays were fabricated at STAR Cryoelectronics and tested at LLNL and FRIB. They have the same high energy resolution of $\sim$1\textendash2~eV in the energy range of interest below 100~eV as before, and 
significantly reduce the presence of both calibration artifacts. We discuss the design changes and the STJ array performance for the next phase of the BeEST experiment.

\end{abstract}


\section{Introduction}
\IEEEPARstart{T}{he} \underline{Be}ryllium \underline{E}lectron capture in \underline{S}uperconducting \underline{T}unnel junctions (BeEST) experiment~\cite{Leach2022} uses $^{\mathrm{7}}$Be-doped superconducting tunnel junction (STJ) sensors for high-resolution recoil spectroscopy of the two-body decay $\mathrm{^7Be}$ + e$^-$~$\mathrm{\rightarrow}$~$\mathrm{^7Li}$~$\mathrm{+}$~$\mathrm{\nu_e}$. The recoil energy of the $\mathrm{^7}$Li daughter is measured accurately and provides a measure of the neutrino mass. Heavy BSM neutrinos would reduce the recoil energy and produce additional peaks in the spectrum as a signature. The BeEST program uses this approach to search for sub-MeV sterile neutrinos~\cite{Friedrich2021}, probe the quantum properties of neutrinos~\cite{Smolsky2025}, and measure recoil dynamics at low energy~\cite{bray-arXiv}. This approach has broad applicability to rare-isotope and fundamental-symmetry studies.
\IEEEpubidadjcol

\IEEEpubidadjcol
\begin{figure}[ht]
	\centering
	\includegraphics[width=0.49\textwidth]{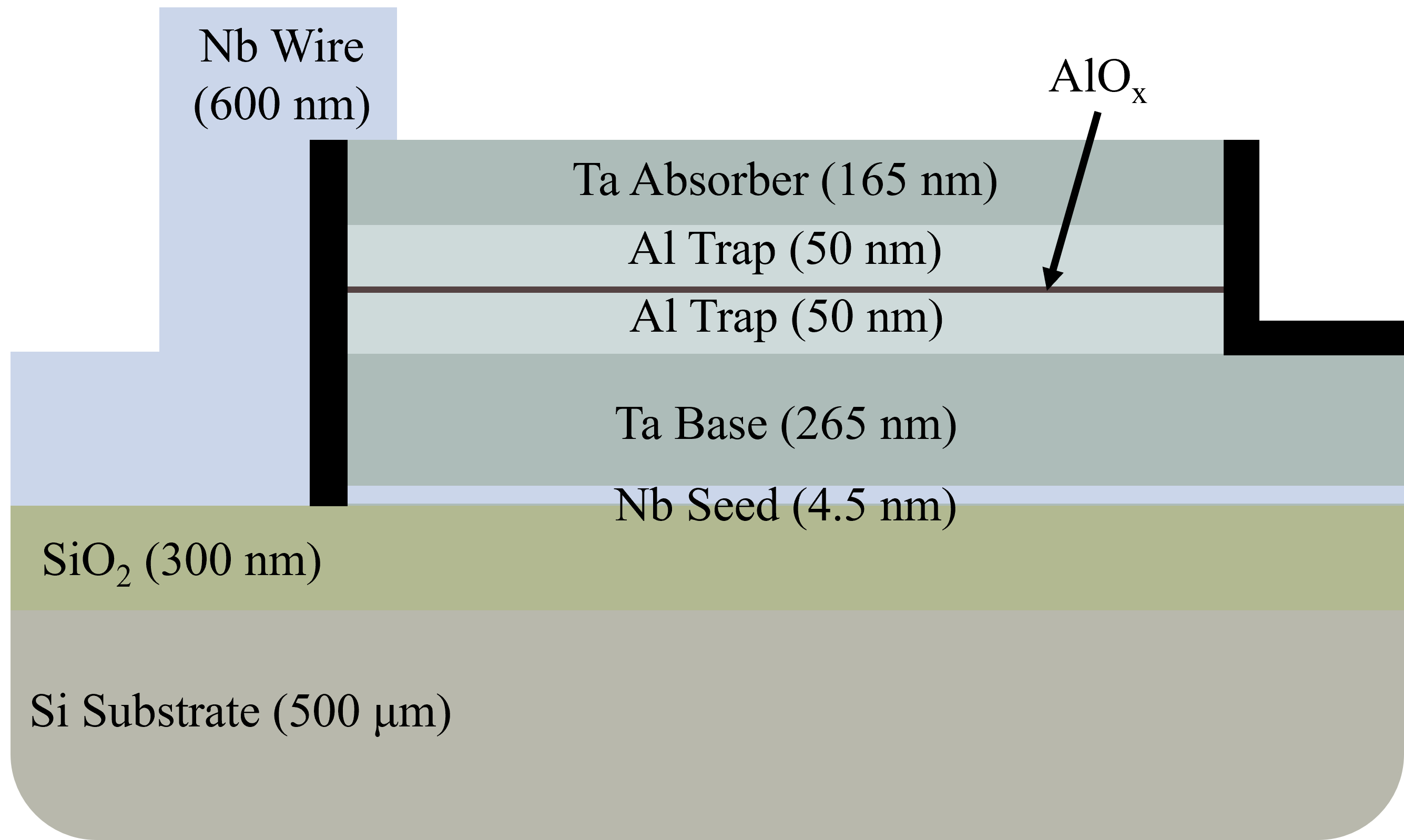}
	\caption{Cross section and film thicknesses for a Ta-Al-AlOx-Al-Ta STJ from STAR Cryoelectronics. The 4.5~nm Nb layer nucleates the base Ta film in the desired bcc \(\mathrm{\alpha}\)-Ta phase.}
	\label{fab-diagram}
\end{figure}

\IEEEpubidadjcol

\begin{figure}[b]
  \centering
  \newlength{\imgsep}\setlength{\imgsep}{0.027\linewidth}
  \begin{minipage}[t]{0.13\linewidth}\centering
    {\normalsize $\mathrm{^{7}}$Be~Decay\strut}\\[0.6ex]
    \includegraphics[width=\linewidth]{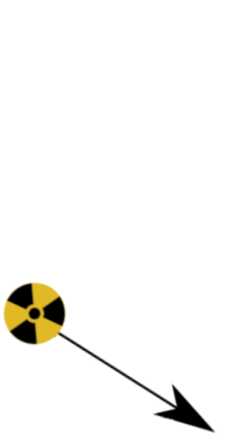}
  \end{minipage}
  \begin{minipage}[t]{0.66\linewidth}\centering
    {\normalsize STJ~Sensor\strut}\\[0.6ex]
    \includegraphics[width=\linewidth]{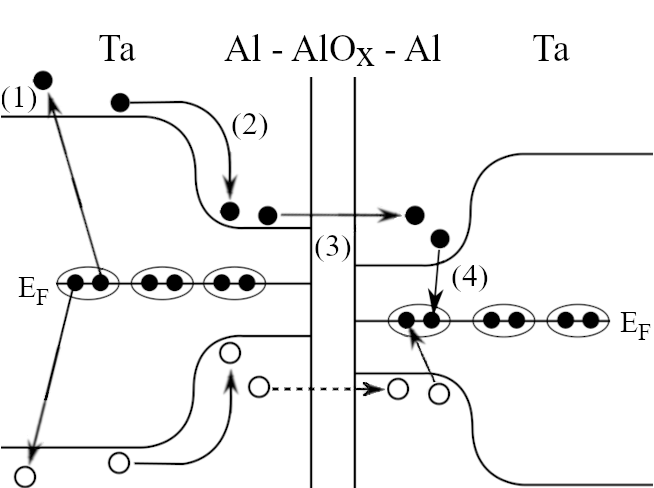}
  \end{minipage}
  \begin{minipage}[t]{0.181\linewidth}\centering
    {\normalsize \hspace{-0.75cm}Signal~Current\strut}\\[0.6ex]
    \includegraphics[width=\linewidth]{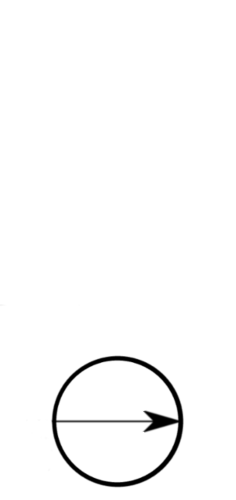}
  \end{minipage}
  \vfill
  \caption{Schematic band diagram and charge flow in an STJ: The EC decay of $^7$Be generates excess quasiparticles in Ta in proportion to the deposited energy~(1), which diffuse into the Al trap and are confined by inelastic scattering~(2). As they tunnel through the barrier~(3), they generate a measurable signal current until they eventually recombine into Cooper pairs~(4). Here, the dashed arrow indicates the direction of negative current flow.}
  \label{Quasi-Tunneling}
\end{figure}

\IEEEpubidadjcol

Superconducting tunnel junctions (STJs) are quantum sensors that consist of two superconducting electrodes separated by a thin insulating barrier. The STJs used by the BeEST experiment follow a 5-layer Ta-Al-AlOx-Al-Ta architecture from STAR Cryoelectronics~\cite{STARCryo} that has provided high energy resolution of $\sim$1~to~$\sim$2~eV FWHM in the past~\cite{Friedrich_2022} (Figure~\ref{fab-diagram}). The top Ta film serves as the absorber for radioactive ions and photons. Its thickness of 165~nm is sufficient to absorb $^{\mathrm{7}}$Be at an implantation energy of 25~keV, and it could be increased in the future for increased efficiency at higher energy. The two Al films are quasiparticle traps that confine the signal charges near the tunnel barrier to increase the tunneling rate~\cite{QPTrapping_Booth}. The Ta base layer ensures a symmetric gap structure in the junction region (Figure~\ref{Quasi-Tunneling}). The base Ta is deposited on a Nb seed layer so that it grows in the desired bcc phase with \(\mathrm{T_c}\)~=~4.5K.

Initially, STJs were developed for astronomy~\cite{Verhoeve_2008} and synchrotron science~\cite{Friedrich_2002, Fons_2006}. For the BeEST experiment, $\mathrm{^7Be}$ decay in the top electrode generates quasiparticles that tunnel across the AlO$_x$ and produce a current pulse with a rise time of a few \(\mathrm{\mu s}\) and a decay time of about \(\mathrm{100~\mu s}\) (Figure~\ref{Quasi-Tunneling}). The pulse is then amplified by a transimpedance preamplifier at room temperature, trapezoidally filtered, and the pulse height is recorded as a measure of the pulse energy~\cite{WARBURTON2015236}.

\begin{figure}[hb]
    \centering
    \includegraphics[width=0.47\textwidth]{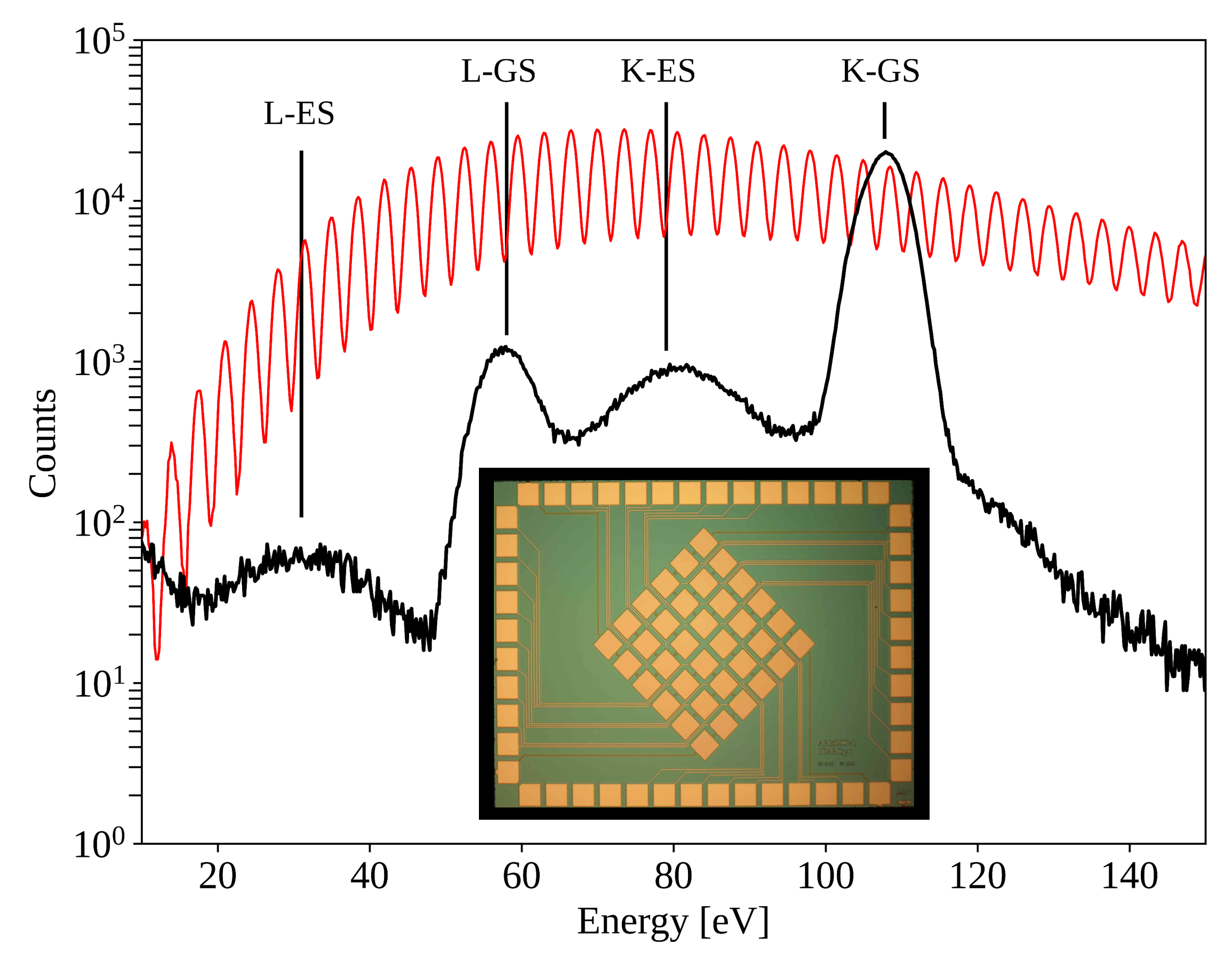}
    \caption{Phase-III $^{\mathrm{7}}$Be decay spectrum from a single STJ pixel (black) and associated laser calibration spectrum (red)~\cite{Inwook}. The four peaks correspond to K-electron capture to the $^{\mathrm{7}}$Li ground state (K-GS), K-electron capture to the $^{\mathrm{7}}$Li excited state (K-ES), and the two corresponding L-capture peaks (L-GS) and (L-ES). The inset shows the 36-pixel array of \((200~\mathrm{\mu m})^2\) STJs used for phase-III measurements.}
	\label{Laser-Spectrum}
\end{figure}

The decay of $^{\mathrm{7}}$Be inside the STJ produces four primary peaks, one for K-capture decay into the ground state of $^{\mathrm{7}}$Li (K-GS), one for decay into the excited state of $^{\mathrm{7}}$Li (K-ES), and the two corresponding L-capture peaks (L-GS and L-ES) (Figure~\ref{Laser-Spectrum})~\cite{Inwook}. Throughout the measurement, the STJ is exposed to a pulsed UV laser that generates a comb of peaks spaced by the single-photon energy of $\sim$3.5~eV. Matching the filtered pulse height with the corresponding photopeak provides in-situ calibration. 
(Figure~\ref{Laser-Spectrum}). 
Accurate STJ calibration is important for the BSM physics goals of the BeEST experiment because the signature of sterile neutrinos consists of an offset spectrum at lower energy~\cite{Friedrich2021}.
In the following sections, we identify artifacts in the calibration data from Phase-III of the BeEST that reduce calibration accuracy. We then demonstrate the 
mitigation of these artifacts by both changing the wiring layout of the STJ arrays and by minimizing the shot-to-shot intensity variation of the pulsed calibration laser.

\begin{figure}[ht]
        \centering
        \includegraphics[width=0.49\textwidth]{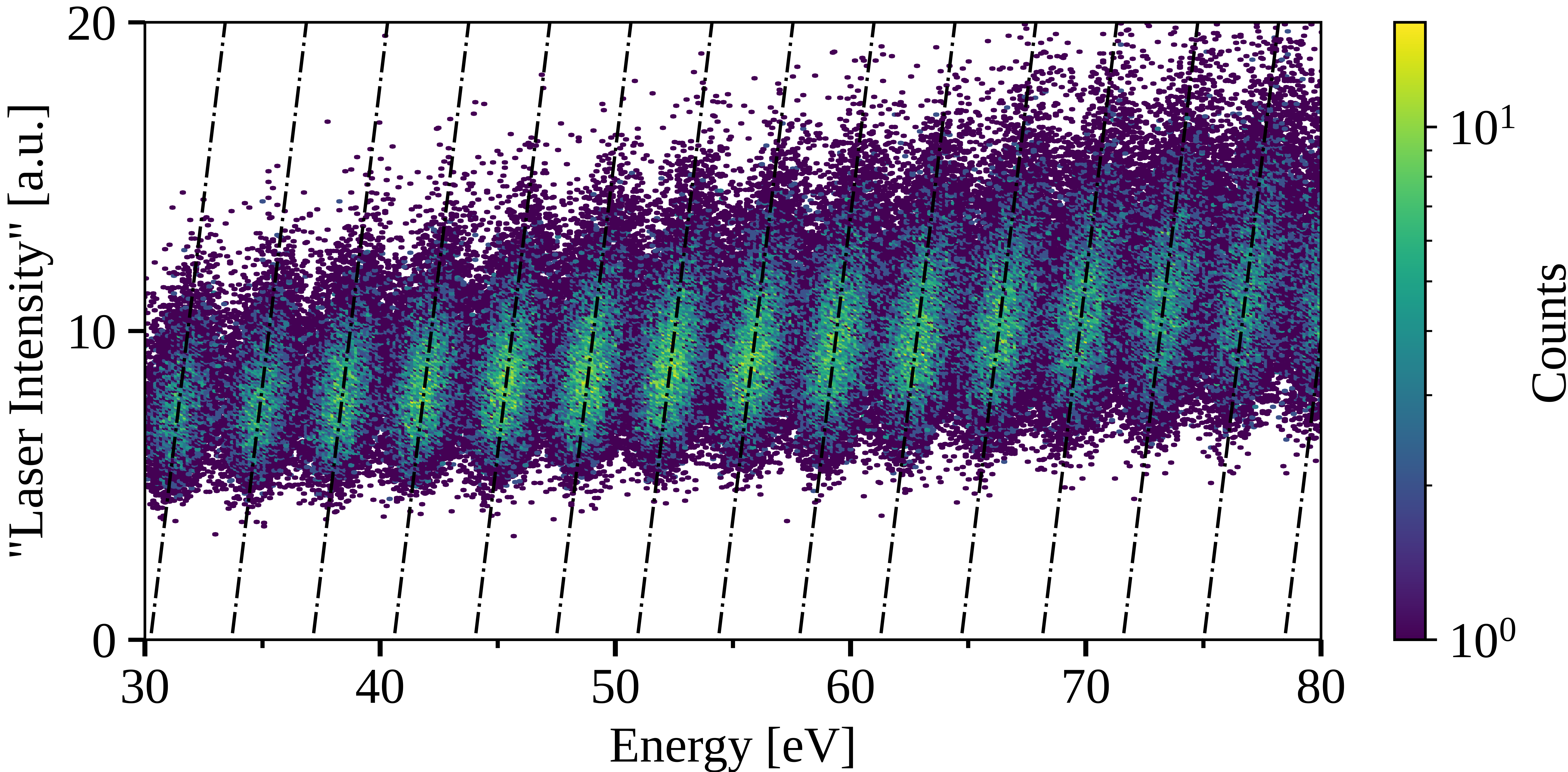}
        \caption{Scatter plot of the signal in one channel vs the laser intensity demonstrating two unexpected artifacts: The clusters are slanted (dashed line), indicating some degree of inter-pixel correlation, and higher-order clusters are offset to a higher average laser intensity~\cite{Inwook}. Here, $\mathrm{E_{\gamma}=3.49865(15)}$~eV~\cite{ponce-thesis} and $\mathrm{FWHM\approx}$ 1.5\textendash \nobreak2.7~eV~\cite{Inwook}.}
        
        \label{Inwook}
    \end{figure}

\section{Systematic Calibration Uncertainties}

Phase-III of the BeEST experiment measured the $^{\mathrm{7}}$Be decay spectrum with a 36-pixel STJ array, in which groups of 9 STJs shared a common ground wire~\cite{Inwook} (Figure~\ref{Laser-Spectrum},~inset).
Throughout the measurement, the STJ response was calibrated with a pulsed 355~nm laser at a rate of 100~Hz. 
We quickly noticed a systematic calibration error because the calibrated $^{\mathrm{7}}$Be spectra from the different pixels did not align.
We later noticed that, for events with a \textit{constant} number of photons, there was a systematic change in the laser signal proportional to laser intensity. 
We can visualize this effect if we take the sum of all simultaneous laser events from all pixels as a measure of the total laser intensity, and plot the signal of one channel against the total laser intensity for each event~(Figure~\ref{Inwook})~\cite{Inwook}. 
The slope of each cluster in the resulting scatter plot shows that events with the \textit{same number} of photons produce systematically larger signals for higher laser intensity.
We realized that this is caused by two effects, namely resistive crosstalk and laser absorption in the Si substrate. The next two sections describe these calibration artifacts and our approach for removing them.


\subsection{Resistive Crosstalk}

\begin{figure}[t]
        \centering
        \includegraphics[width=0.49\textwidth]{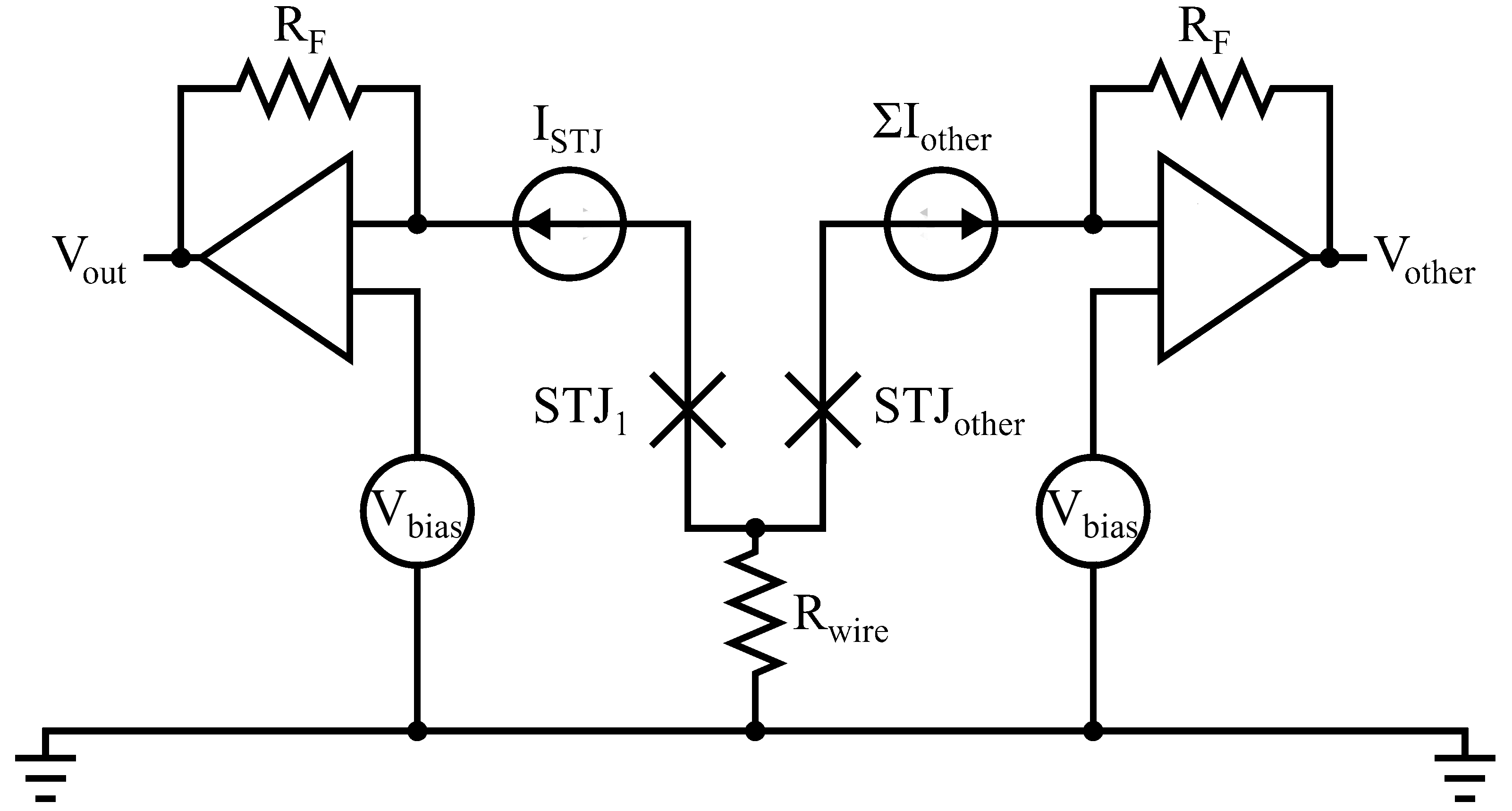}
        \caption{Schematic of the readout electronics for the calibration of two STJs with a common ground wire. Here, STJ$_{\mathrm{other}}$ contributes to the current in the resistive ground plane for simultaneous laser events as a result of 
        $
       \sum \mathrm{ I_{other}}
        $ 
        in Equation 1. However, this is not the case for randomly occurring $^{\mathrm{7}}$Be decay events.}
    \label{Crosstalk}
\end{figure}

Resistive crosstalk occurs as a result of the common ground wire shared between multiple STJs, which has a resistance R$_{\mathrm{wire}}$ between the STJ and the preamplifier. Each signal I$_{\mathrm{signal}}$ produces a small voltage drop I$_{\mathrm{signal}}$R$_{\mathrm{wire}}$ across this resistance, which is then amplified by (R$_{\mathrm{F}}$~/~R$_{\mathrm{STJ}}$) in all STJs that share the same ground wire (Figure~\ref{Crosstalk}). 
This is not a problem if signals occur randomly in time~\cite{STJs-Carpenter}, as they do in the decay of $^{\mathrm{7}}$Be. The calibration signals, however, occur simultaneously in all pixels for each laser pulse.  This adds a current in each STJ that depends on the sum of all simultaneous currents in other STJs (\(\mathrm{\sum I_{other}}\)) that share the same ground wire. It changes the output to

\begin{equation}
    V_{out}=R_{F}I_{STJ}\bigg{(}1+\frac{\sum I_{other}}{I_{STJ}}\frac{R_{wire}}{R_{STJ}}\bigg{)}
    \label{Vout-2}.
\end{equation}

Resistive crosstalk depends not only on the resistance of the shared ground wire, but also on the dynamic resistance (\(\mathrm{R_{STJ}}\)) of the STJ at the bias point. Crosstalk is one of the contributions to the slope of the scatter plots in Figure~\ref{Inwook} and the systematic variation in the laser calibration for different pixels. While it can be corrected based on the measured slope, it increases the calibration uncertainty to 20~meV~\cite{Inwook}.
This motivates the redesign of our STJ arrays with separate ground wires per pixel for phase-IV of the BeEST experiment.

\subsection{Substrate Events}

If photons from the calibration laser are absorbed in the Si substrate between pixels, they produce athermal phonons that can diffuse to the bottom STJ electrode before thermalizing and break Cooper pairs. This produces excess quasiparticles that contribute an offset to $\mathrm{I_{STJ}}$ that depends on the number of substrate photons and their absorption location. 
To reduce the number of substrate photons, a Si collimator is placed $\sim$100~$\mathrm{\mu m}$ in front of the STJ sensors to restrict the illumination to the active pixel area. However, scattering still causes some of them to be absorbed in the Si substrate between pixels.
If the laser intensity is constant, there will only be statistical fluctuations in the number of substrate events whose average is constant and can be readily subtracted from the data~\cite{Ponce_2016}.

In the BeEST experiment, the calibration laser is attenuated so that it produces a comb of peaks in the region of interest from $\sim$20 to $\sim$120~eV. In phase-III, the laser was attenuated by reducing the pump current. However, this introduced an unintentional shot-to-shot variation of the laser intensity, as the pump current was outside the recommended operating range.
This produced a calibration spectrum that is broader than a Poissonian distribution (Figure~\ref{Laser-Spectrum}). It also produces systematic variations in the number of substrate events, and therefore a systematic change in offset as a function of laser intensity. This mimics a gain change of the laser signal relative to the $^{\mathrm{7}}$Be signal.

The shot-to-shot intensity variations of the laser also contribute to the slope in the scatter plot in Figure~\ref{Inwook}. Laser pulses with higher intensity will produce a distribution of slanted clusters with a higher average number of substrate photons, and thus a higher average offset.
While we can correct for this effect in our calibration~\cite{Inwook}, it reduces the calibration accuracy.
We therefore no longer reduce the pump current to adjust the laser output. Rather, we operate the laser at full power and adjust its output with a mechanical attenuator that transmits the laser light across an air gap with an adjustable width. At FRIB, we also now use a 261.8(5)~nm laser with a single-photon energy of 4.736(9)~eV to increase the spacing between calibration peaks.

\section{STJ Sensor Design}

The new wafer contains STJ arrays of 32, 64, and 128 pixels that are matched to the 32-channel preamplifier card from XIA LLC~\cite{WARBURTON2015236}. The pixels have areas of (70~$\mathrm{\mu m}$)$^{\mathrm{2}}$, (130~$\mathrm{\mu m}$)$^{\mathrm{2}}$ and (200~$\mathrm{\mu m}$)$^{\mathrm{2}}$ so that arrays can be chosen for high energy resolution or large active volume depending on the application~\cite{STJs-Carpenter}.
The STJs are rotated by 45 degrees with respect to the chip edges to efficiently suppress the Josephson current in all pixels~\cite{Peterson_1991}. 

\begin{figure}[hb]
  \centering
  \newlength{\imgsepy}
  \setlength{\imgsepy}{0.027\linewidth} 

  \begin{minipage}[t]{0.47\linewidth}\centering
    \includegraphics[width=\linewidth]{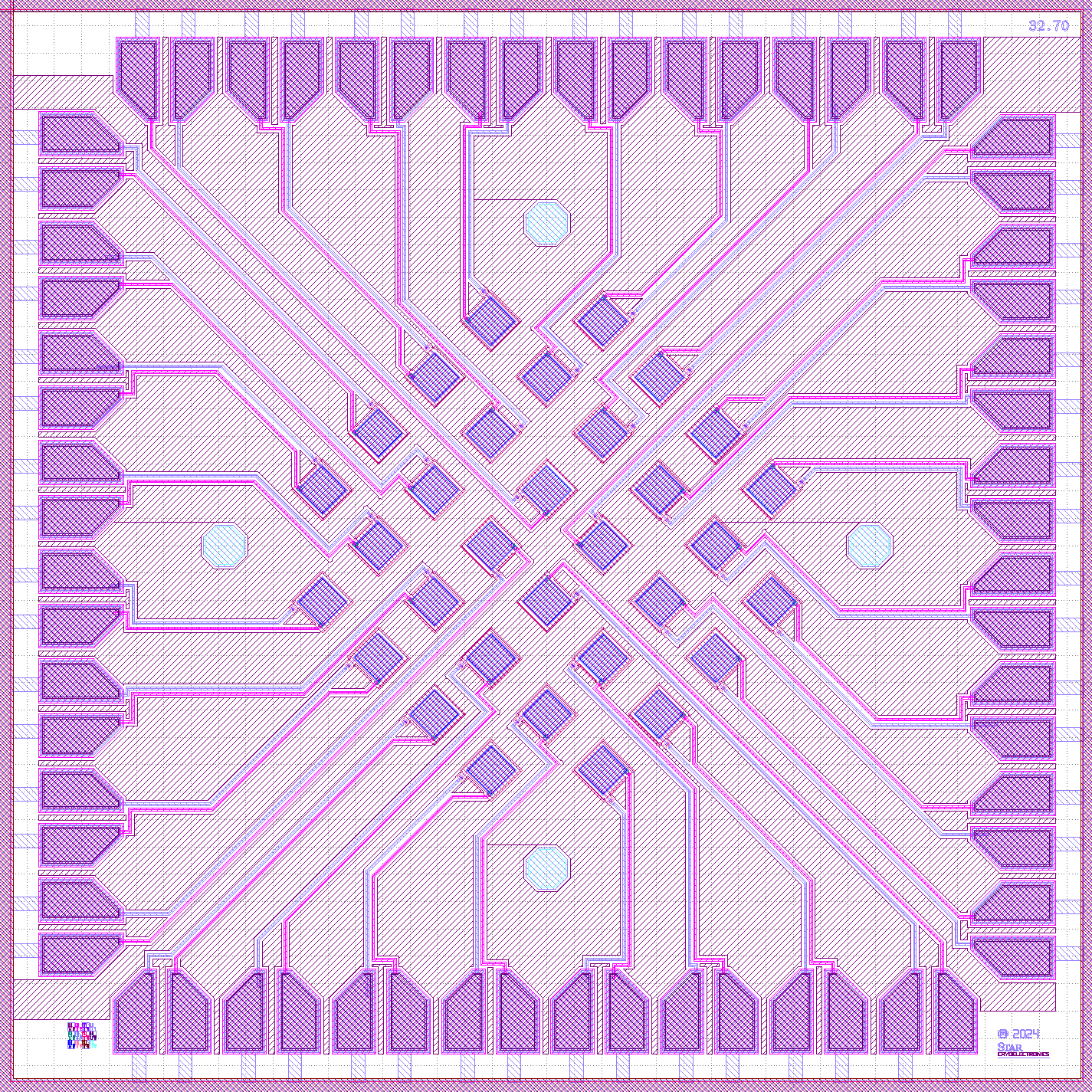}
  \end{minipage}
  \hspace{\imgsepy}
  \begin{minipage}[t]{0.47\linewidth}\centering
    \includegraphics[width=\linewidth]{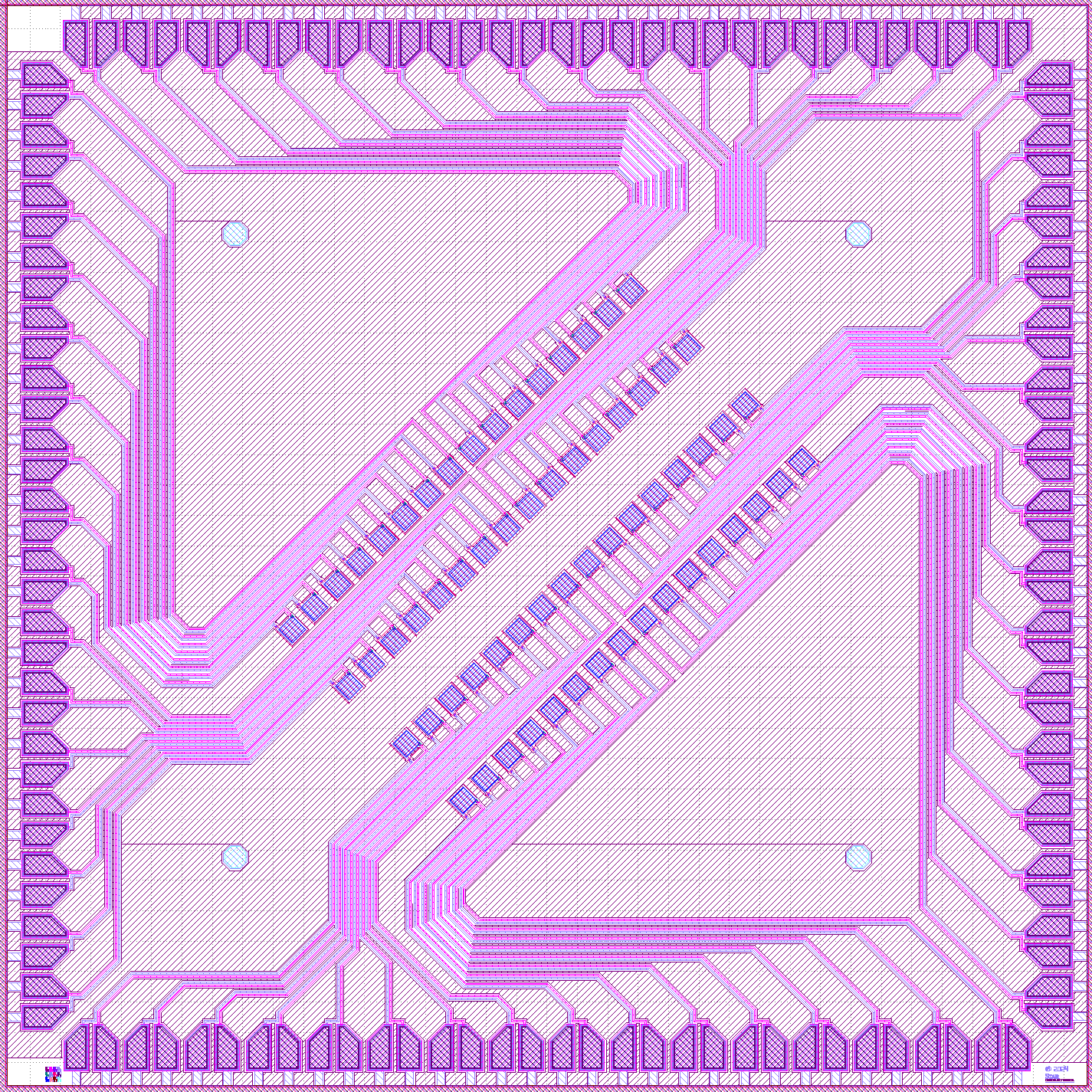}
  \end{minipage}

  \vspace{1.8\imgsepy} 

  \begin{minipage}[t]{0.47\linewidth}\centering
    \includegraphics[width=\linewidth]{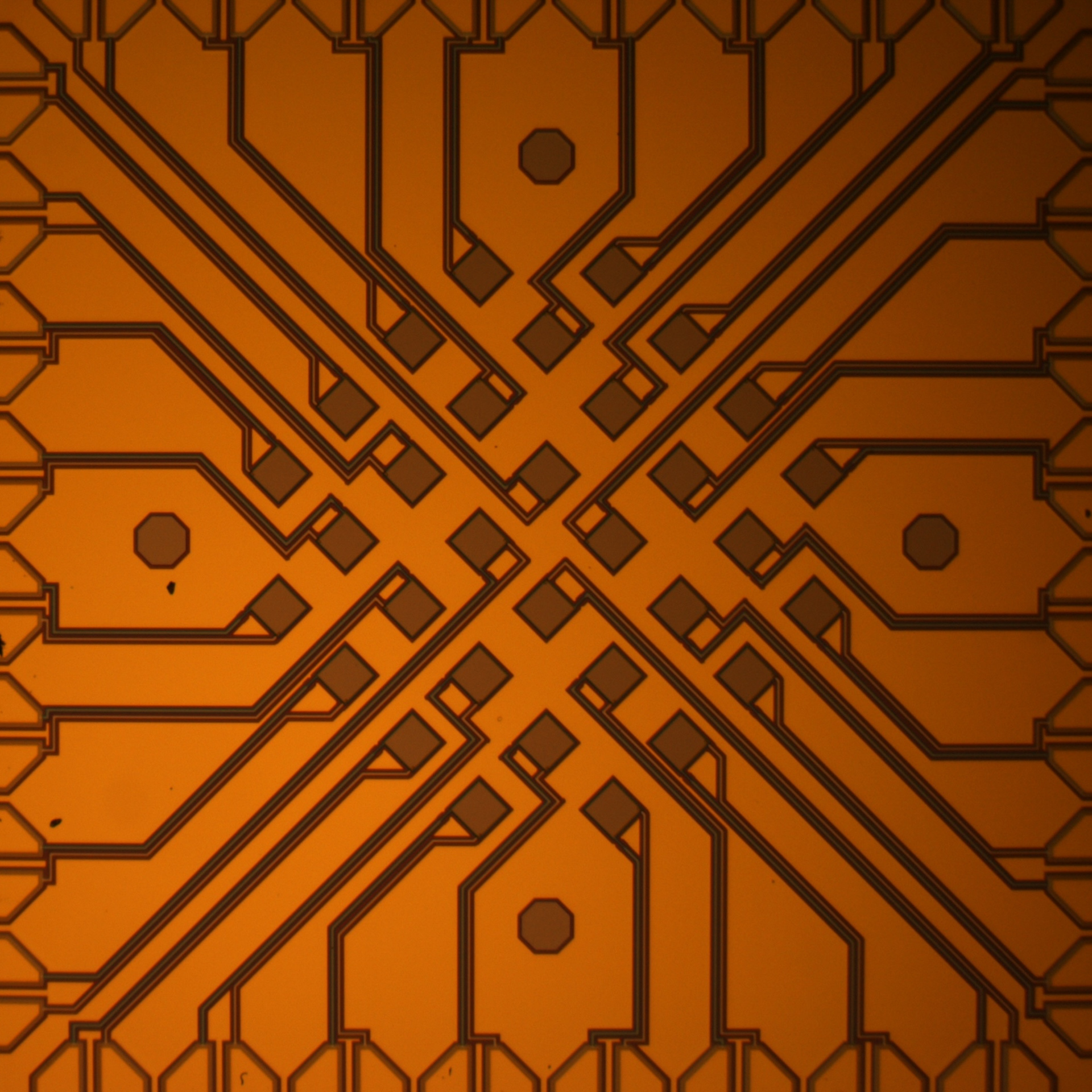}
  \end{minipage}
  \hspace{\imgsepy}
  \begin{minipage}[t]{0.47\linewidth}\centering
    \includegraphics[width=\linewidth,height=\linewidth]{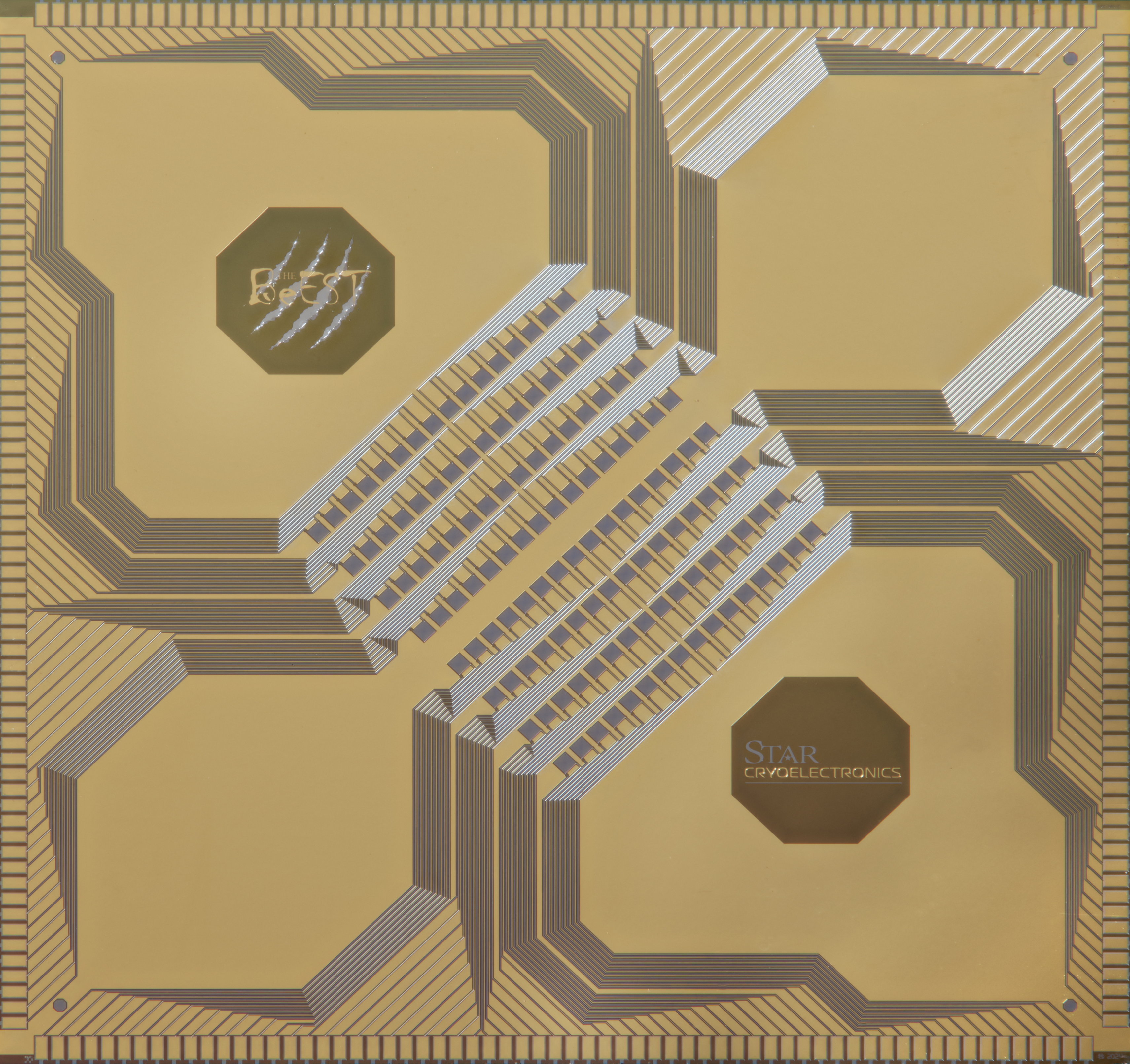}
  \end{minipage}

  \caption{Top: KLayout designs of a 32-pixel (left) and a 64-pixel (right) array of (130$\mathrm{\mu m}$)$^{\mathrm{2}}$ STJs. Bottom: Photographs of a 32-pixel (left) and a 128-pixel (right) sensor array. The 32-pixel array was used for the spectra in Figure~\ref{Results-Spectra}.}
  \label{Sensors}
\end{figure}

The main difference to previous designs is that each STJ now has its own ground wire to eliminate resistive crosstalk (Section II A). 
Each pixel is connected to a twisted pair of wires in a differential pair configuration that is matched to the input of the preamplifier electronics. 
As a consequence, the sensor arrays now require almost double the number of bond pads of previous designs. 
These bond pads have a 200~$\mathrm{\mu m}$ pitch, which increases the chips sizes to (4~mm)$^{\mathrm{2}}$, (7.2~mm)$^{\mathrm{2}}$, and (13.8~mm)$^{\mathrm{2}}$ for the 32-, 64- and 128-pixel arrays, respectively (Figure~\ref{Sensors}). The new sensor array chips also incorporate a gold thermalization layer to absorb and downconvert phonons from substrate events. The chips were designed using \textit{KLayout}~\cite{KLayout} and fabricated at STAR Cryoelectronics~\cite{STARCryo}.


\section{Results}

A 32-pixel array of (130~$\mathrm{\mu m}$)$^{\mathrm{2}}$ STJs was tested at Lawrence Livermore National Laboratory (LLNL) in a ``wet" adiabatic demagnetization refrigerator (ADR) with liquid N$\mathrm{_2}$ and He pre-cooling. The STJs were operated at a temperature of $\sim$100~mK and biased at 100~$\mathrm{\mu V}$. 
They had a leakage current of $\sim$10~nA and a dynamic resistance of 10~\(\mathrm{k\Omega}\), comparable to STJs from earlier fabrication runs. They were exposed to the same pulsed \(\mathrm{355}\)~nm laser used in phase-III at a rate of 2~kHz. 
Since the STJ pitch in the arrays from the new chip is slightly different, the older Si collimators were no longer well-matched. Only 5 of the 32 pixels were fully exposed to the laser, while others were partially or completely covered by the collimator. 

\begin{figure}[t]
        \centering
        \hspace{-0.3cm}
        \includegraphics[width=0.49\textwidth]{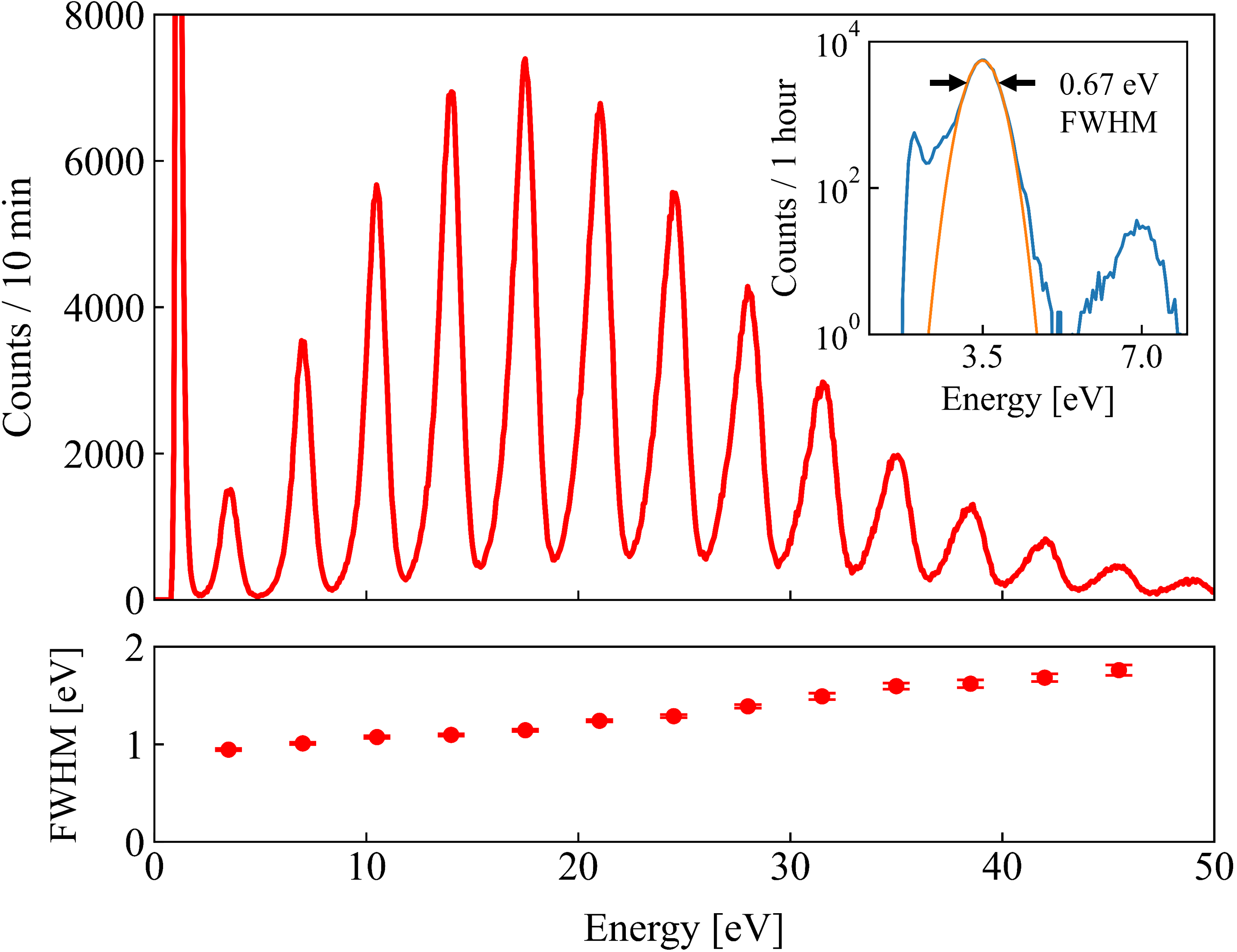}
        \caption{(Top): Laser calibration spectrum from one STJ of the 32-pixel array at LLNL. Below 40~eV, the calibration precision was 21~meV.
        (Bottom): The FWHM energy resolution increases as a function of energy due to statistics as expected. The inset shows the spectrum of a pixel that was only indirectly illuminated.}
        \label{Results-Spectra}
    \end{figure}

The laser spectrum from one of the fully exposed STJs is shown in Figure~\ref{Results-Spectra}. At low energies, the resolution is limited to $\sim$1~eV FWHM by electronic noise. It increases to $\sim$2~eV at 50~eV as the statistical STJ noise increases (Figure~\ref{Results-Spectra}, Bottom).
This performance is similar to earlier STJs and illustrates the reproducibility of the STJ fabrication process at STAR Cryoelectronics.

Interestingly, one of the pixels that was completely covered by the Si collimator had an energy resolution of 0.67~eV FWHM at 3.5~eV (Figure~\ref{Results-Spectra},~top~right). This STJ was only exposed to scattered photons, and the intensity was therefore significantly lower, with only the single-photon peak recording a significant number of events. Still, a resolution of 0.67~eV FWHM is the highest resolution we have measured in this array. Since the leakage current of this STJ was similar to that of other pixels in this array, the electronic noise that dominates the energy resolution at 3.5~eV should have been similar, too.
This suggests that some of the limiting noise is not just due to the readout electronics but also due to variations in the number of substrate photons absorbed in the vicinity of an STJ. Future experiments should try to better collimate the calibration laser to reduce this effect. 

A 64-pixel array of (130~$\mathrm{\mu m}$)$^{\mathrm{2}}$ STJs from the same wafer was subsequently tested in a ``dry" ADR with pulse-tube pre-cooling and a base temperature of 40~mK.
The tests were part of the initial installation of the ADR for the Superconducting Array for Low Energy Radiation (SALER) experiment at the Facility for Rare Isotope Beams (FRIB)~\cite{Drew-Talk}. The STJ array was operated without a Si collimator at the end of a cold finger, with a hole in the magnetic shielding to implant radioactive ions into the STJ. 

At FRIB, the vibrational noise from SALER's dry ADR, as well as magnetic and RF noise from nearby equipment, make this environment noisier than the experimental setup at LLNL. 
The STJs had an average dynamic resistance 
of $\sim$2~$\mathrm{k\Omega}$ and leakage currents of $\mathrm{\lesssim100~nA}$.
The high leakage current and relatively low dynamic resistance suggest that the STJs had trapped some flux and that the magnetic shielding at FRIB could be improved. 
The STJs were biased between $\sim$75\nobreak \textendash \nobreak100~$\mathrm{\mu V}$, with leakier junctions requiring a lower bias voltage for best performance.
The STJs were exposed to a pulsed {262}~nm laser from CrystaLaser~\cite{Crystalaser} at a rate of 2~kHz. The laser was operated at full power and attenuated with a variable mechanical attenuator to minimize fluctuations in intensity.

\begin{figure}[t]
    \centering
    \begin{minipage}[t]{1\linewidth}\centering
        \hspace{-0.15cm}
        \includegraphics[width=1\textwidth]{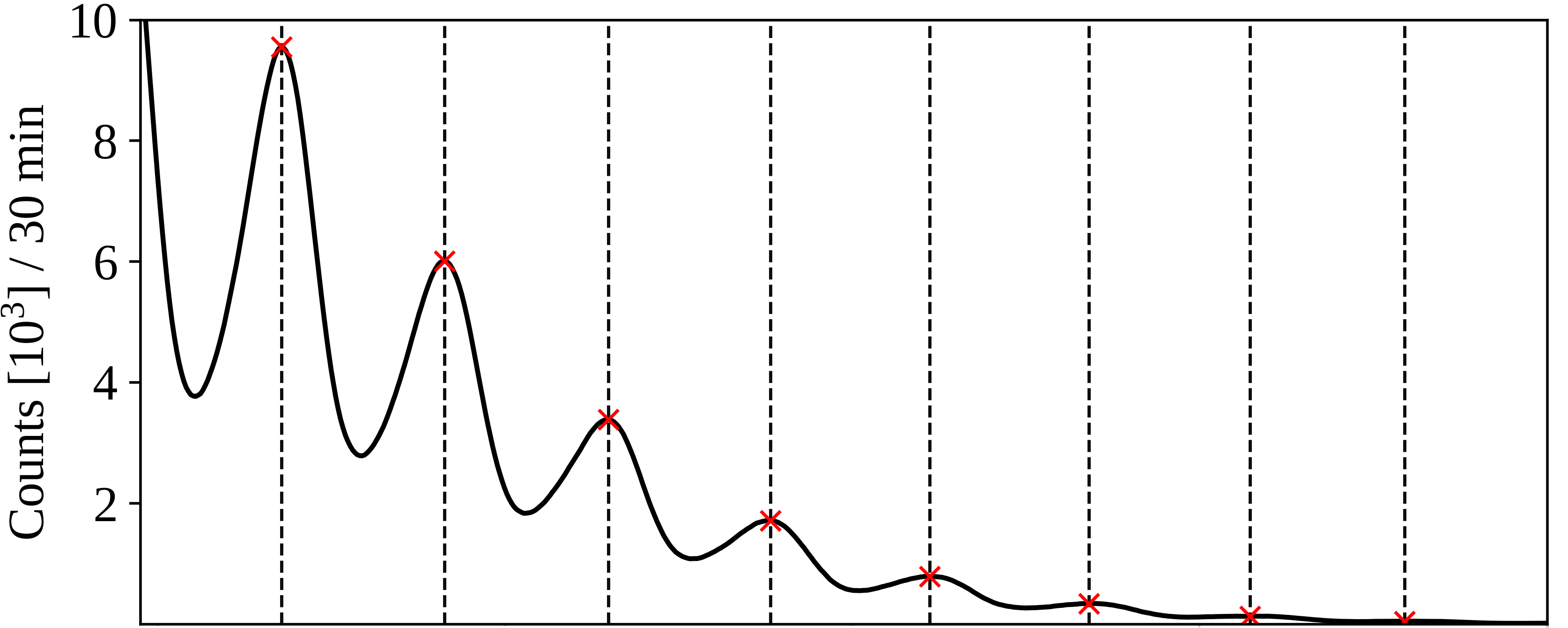}
    \end{minipage}
    \vspace{0.2cm}
    \begin{minipage}[t]{1\linewidth}
    \centering
        \includegraphics[width=1\textwidth]{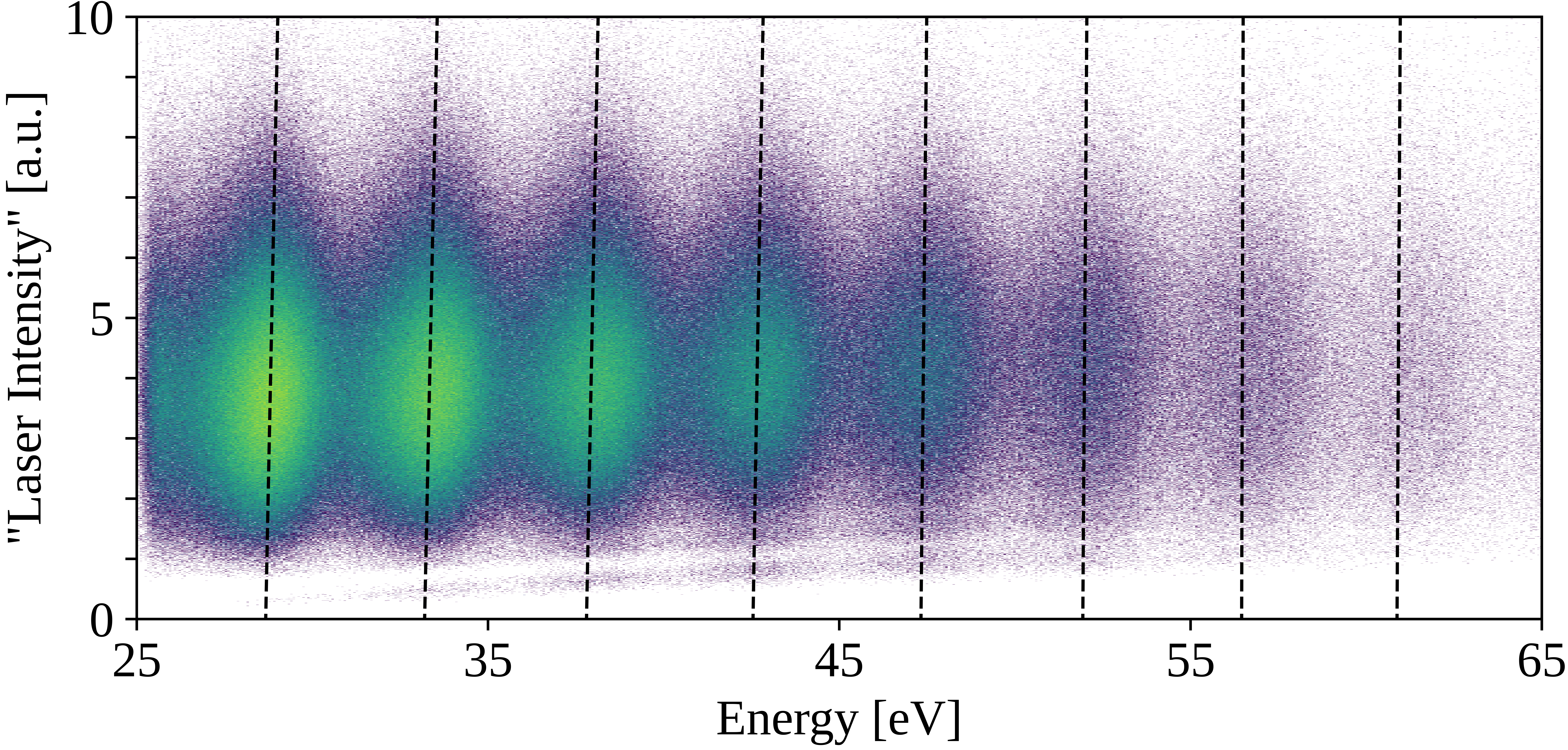}
    \end{minipage}
    \caption{Laser spectrum (Top) and intensity-dependent calibration results (Bottom) of differential pair STJ array. Data was taken at FRIB with a 64-STJ array. These results demonstrate the significant reduction of our phase-III calibration artifacts: cluster slope and cluster offset. Here, $\mathrm{E_{\gamma} \approxeq 4.736(9)}$~eV and the FWHM is $\sim$1.9\textendash 3.6~eV in the energy range below \(\mathrm{50~eV}\). Below \(\mathrm{40~eV}\), the calibration precision is \(\mathrm{17~meV}\). The horizontal band at the bottom of the plot is a triggering artifact.} 
    \label{Results}
\end{figure}


    
Under these conditions, the STJs had an energy resolution of $\sim$1.9 to $\sim$3.6~eV FWHM in the energy range below 50~eV (Figure~\ref{Results}, Top). This is only slightly worse than in the earlier experiment at LLNL, where the STJs were operated inside a well-shielded, wet cryostat. It is consistent with the increased electronic noise due to the higher leakage currents and the reduced dynamic resistance of the STJ at the bias point.

In order to determine by what factor we have reduced the calibration artifacts, we again plot the sum of simultaneous STJ signals as a measure of the laser intensity against the response of a single STJ signal (Figure~\ref{Results}, Bottom). Calculating the slope of all clusters, we find that the separate-ground array reduces the artifact by a factor of 2.7, which will improve the accuracy of the calibration but indicates some residual effect likely from substrate phonons. This is confirmed by calculating the correlation between one pixel's energy and the laser intensity, which reduced from $0.72$ in phase-III data to $0.19$ in the separate-ground arrays. Complete elimination of this artifact will likely require an intensity-stabilized laser, but these results already demonstrate a large reduction in both effects. 

\section{Summary}

We are developing superconducting tunnel junction (STJ) sensor arrays to search for physics beyond the standard model (BSM) with accurate measurements of nuclear decays.
We have identified two systematic errors in the laser calibration of our STJ detectors. Both are due to the fact that nuclear decays occur randomly in time, while the photons from a pulsed calibration laser arrive in all STJ pixels simultaneously.
One error is due to resistive crosstalk between pixels, because multiple pixels used to share a single ground in earlier STJ arrays to reduce the number of wires to the cryostat cold stage. This changes the laser signal of each STJ detector in proportion to the simultaneous signal currents of all other STJs that share the same ground wire.
The other error is caused by absorption of laser photons in the Si substrate \textit{if} the laser intensity fluctuates significantly from shot-to-shot. Since substrate photons cause a signal offset, fluctuations in laser intensity change the effective gain of the laser calibration relative to the signal from nuclear decays. While both effects can be corrected for after data acquisition, they reduce the calibration accuracy.

We have therefore redesigned the STJ arrays such that each pixel now has its own ground wire. The new STJs have an energy resolution between 1\textendash 2~eV FWHM in the energy range of interest below 100~eV, similar to earlier STJs fabricated with the same process parameters. 
We are also now operating the calibration laser under conditions where its output fluctuates significantly less. We have shown that these two changes significantly reduce both the presence of calibration artifacts and the dependence of the calibration on the laser intensity. 
Interestingly, one pixel that was only illuminated very weakly by scattered laser light had an exceptionally good energy resolution of 0.67~eV at 3.5~eV. Since this pixel was also affected less by laser photons in the surrounding substrate, this suggests that substrate phonons do not just affect the responsivity, but also the noise of an STJ.
Better collimation of the laser to the active area of the STJs may therefore further improve their energy resolution.
The new STJ arrays are well-suited for phase-IV of the BeEST experiment and for other BSM searches such as SALER at FRIB.

\section*{Acknowledgments}

This work is funded by the Gordon and Betty Moore Foundation (no. 10.37807/GBMF11571), the US DOE Office of Nuclear Physics awards DE-SC0021245 and SCW1758, the LLNL LDRD program grant 20-LW-006, and the DOE SBIR grant DE-SC0024810. TRIUMF receives federal funding through a contribution agreement with the National Research Council of Canada. This work was performed under the auspices of the US Department of Energy by LLNL under contract no. DE-AC52-07NA27344. F.P. is funded as part of the Open Call Initiative at PNNL and conducted under the LDRD program. PNNL is a multiprogram national laboratory operated by Battelle for the US Department of Energy.

\bibliographystyle{IEEEtran}
\bibliography{LTD_2025}

\newpage

\vfill
\end{spacing}
\end{document}